# Giant magnetoresistance in electrodeposited Co-Cu/Cu multilayers: origin of absence of oscillatory behaviour

I. Bakonyi*, E. Simon, B.G. Tóth, L. Péter and L.F. Kiss

Research Institute for Solid State Physics and Optics, Hungarian Academy of Sciences.

H-1525 Budapest, P.O.B. 49, Hungary

*Keywords: giant magnetoresistance (GMR); oscillatory coupling; electrodeposited multilayers, electrical resistivity; coercive force*

PACS numbers: 75.47.De; 75.70.-i; 75.70.Cn; 81.15.Pq

**Abstract** ─ A detailed study of the evolution of the magnetoresistance was performed on electrodeposited Co/Cu multilayers with Cu layer thicknesses ranging from 0.5 nm to 4.5 nm. For thin Cu layers (up to 1.5 nm), anisotropic magnetoresistance (AMR) was observed whereas multilayers with thicker Cu layers exhibited clear giant magnetoresistance (GMR) behaviour. The GMR magnitude increased up to about 3.5 to 4 nm Cu layer thickness and slightly decreased afterwards. According to magnetic measurements, all samples exhibited ferromagnetic (FM) behaviour. The relative remanence turned out to be about 0.75 for both AMR and GMR type multilayers. This clearly indicates the absence of an antiferromagnetic (AF) coupling between adjacent magnetic layers for Cu layers even above 1.5 nm where the GMR effect occurs. The AMR behaviour at low spacer thicknesses indicates the presence of strong FM coupling (due to, e.g., pin-holes in the spacer and/or areas of the Cu layer where the layer thickness is very small). With increasing spacer thickness, the pin-hole density reduces and/or the layer thickness uniformity improves which both lead to a weakening of the FM coupling. This improvement in multilayer structure quality results in a better separation of magnetic layers and the weaker coupling (or complete absence of interlayer coupling) enables a more random magnetization orientation of adjacent layers, all this leading to an increase of the GMR. Coercive field and zero-field resistivity measurements as well as the results of a structural study reported earlier on the same multilayers provide independent evidence for the microstructural features established here. A critical analysis of former results on electrodeposited Co/Cu multilayers suggests the absence of an oscillating GMR in these systems. It is pointed out that the large GMR reported previously on such Co/Cu multilayers at Cu layer thicknesses around 1 nm can be attributed to the presence of a fairly large superparamagnetic (SPM) fraction rather than being due to a strong AF coupling. In the absence of SPM regions as in the present study, AMR only occurs at low spacer thicknesses due to the dominating FM coupling.

---

*Corresponding author. E-mail: bakonyi@szfki.hu

# I. INTRODUCTION

Soon after the discovery of the giant magnetoresistance (GMR) effect in layered magnetic nanostructures,[1,2] it was shown that in magnetic/non-magnetic multilayers the GMR magnitude oscillates with the thickness of the non-magnetic (NM) spacer layer.[3] This has been demonstrated for many multilayer systems prepared by physical deposition methods like sputtering, evaporation or molecular-beam epitaxy (MBE). The oscillatory behaviour finds its natural explanation in the corresponding oscillations of the sign of the exchange coupling of adjacent layer magnetizations.[3-5] Namely, for spacer layer thicknesses yielding antiferromagnetic (AF) coupling (for Co/Cu multilayers this occurs at about 1, 2 and 3 nm Cu layer thicknesses), the adjacent layer magnetizations have an antiparallel alignment in zero external magnetic field, which state is accompanied by a high electrical resistance. By applying a sufficiently large magnetic field, all the layer magnetizations are aligned parallel, which state has a lower resistance than the zero-field value and this yields a GMR effect. For spacer layer thicknesses resulting in a ferromagnetic (FM) coupling between adjacent magnetic layers, there is no change in the magnetization alignments upon the application of an external field and the GMR effect does not occur (in such cases, just the conventional anisotropic magnetoresistance (AMR) of bulk ferromagnets[6] can be observed).

By contrast, whereas a significant GMR effect was demonstrated also for electrodeposited multilayers already 15 years ago,[7] reports concerning an oscillatory GMR behaviour on such systems have remained fully controversial till today, in spite of the extensive research works in this area (see the reviews in Refs. 8 and 9). Two or more peaks in the spacer layer thickness dependence of the GMR magnitude have been reported for electrodeposited multilayers such as Ni-Cu/Cu,[10,11] Co-Cu/Cu,[11-15] Co-Ni-Cu/Cu,[16-18] Fe-Co-Cu/Cu[19] And Co-Ag/Ag[12] (due to the commonly applied single-bath technique,[8,9] the magnetic layer of electrodeposited multilayers unavoidably contains a few percent of the non-magnetic element). These peaks were often claimed as resulting from an oscillatory exchange coupling between the adjacent layer magnetizations. It should be noted, however, that the position, separation and relative amplitude of these peaks in most cases did not correspond to the relevant values obtained on physically deposited multilayers of related compositions. On the other hand, an initial monotonic increase of GMR magnitude which then eventually flattened off or, after a single maximum, decreased for higher spacer layer thicknesses was reported for electrodeposited multilayers such as Ni-Cu/Cu,[20,21] Co-Cu/Cu,[18,22-28] Co-Ni-Cu/Cu,[29-33] Fe-Co-Ni-Cu/Cu,[34,35] Co-Au/Au,[36] Co-Ag/Ag[36] and for an electrodeposited spin-valve system with alternating hard and soft magnetic layers $Ni_{93}Fe_4Cu_3/Cu/Ni_{78}Fe_{14}Cu_8/Cu$.[37] The appearance or absence of a plateau or a maximum was dependent mainly on the maximum spacer layer thickness investigated and the position of the plateau region or the maximum varied from study to study, the maximum position being at around 1 to 2 nm for Ni-Cu/Cu, Co-Au/Au and Co-Ag/Ag and around 3 to 6 nm for Co-Cu/Cu and Co-Ni-Cu/Cu. Even a monotonic decrease of the GMR magnitude with Cu layer thickness, with a levelling off at around 2 nm, was reported for Co-Ni-Cu/Cu multilayers.[38]

On the other hand, there has been recently a significant progress in understanding the electrochemical processes governing deposit formation[21,25,26,39-42] and the factors influencing the GMR characteristics, with special reference to the appearance of a possible superparamagnetic (SPM) contribution to the total GMR[26,28,41-46] in electrodeposited multilayers. This instigated us to undertake a thorough study of the evolution of the GMR magnitude in electrodeposited Co/Cu multilayers with Cu layer thicknesses from 0.5 nm to 4.5 nm in steps of about 0.1 nm. It was expected that these new results on multilayers prepared under carefully controlled conditions[40,42] might help to resolve these long-lasting controversies. Magnetic hysteresis loops and zero-field resistivities were also measured in order to get additional data for characterizing these multilayers. A structural study on the same multilayer series has been reported recently.[47]

The results of the present study revealed a systematic evolution of the MR behaviour with copper layer thickness. Below and above about 1.5 nm Cu layer thickness, AMR and GMR behaviour, respectively, can be observed. Beyond this critical thickness, the GMR magnitude shows a monotonic increase of up to about 3.5 to 4 nm Cu layer thickness and a slight decrease afterwards. Such a variation of the MR behaviour can be conclusively explained by assuming the presence of a large density of pin-holes in the spacer and/or fairly strong spacer thickness fluctuations for Cu layer thicknesses below 1.5 nm and an improved continuity and thickness uniformity of the Cu layers above 1.5 nm thickness. The drop of the zero-field resistivity and the bulk-like low coercive field for the smallest Cu layer thicknesses give further support for the presence of pin-holes as one definitive cause of the observed AMR behaviour. The diminution of the zero-field resistivity and the increasing coercivity for larger Cu layer thicknesses, on the other hand, indicate that the magnetic layers become more and more efficiently separated as the Cu layer thickness gets sufficiently large and the Cu thickness uniformity also improves. Due to the reduction of FM coupling, the magnetic layers become progressively uncoupled and their random magnetization orientation can then give rise to an increasingly larger GMR effect as observed experimentally. The structural results reported separately for the same multilayers[47] well corroborate the microstructural features deduced here.

The paper is organized as follows. In Section II, the sample preparation and characterization as well as the measurement techniques are briefly described. The results of various measurements on a Co/Cu multilayer series with varying spacer layer thickness are presented in Section III. Section IV provides a discussion of the results and a comparison with findings of former investigations. Finally, Section V summarizes the main conclusions of this study.

## II. EXPERIMENTAL DETAILS
### A. Sample preparation and characterization

An aqueous electrolyte containing 0.8 mol/ℓ $CoSO_4$, 0.015 mol/ℓ $CuSO_4$, 0.2 mol/ℓ $H_3BO_3$ and 0.2 mol/ℓ $(NH_4)_2SO_4$ was used to prepare magnetic/non-magnetic Co/Cu multilayers by using a G/P pulse combination[39] in which a galvanostatic (G) and a potentiostatic (P) pulse is applied for the deposition of the magnetic and the non-magnetic layer, respectively. The Cu deposition potential was optimized according to the method described in Ref. 40 which ensured that neither Co dissolution nor Co codeposition occurred during the Cu deposition pulse. Under the conditions applied, the Cu-incorporation into the magnetic layer is fairly low [the composition is approximately $Co_{99.4}Cu_{0.6}$ (see Ref. 47)] and this justifies referring to the magnetic layer of our samples as a Co layer. The electrodeposition was performed in a tubular cell of 8 mm x 20 mm cross section with an upward looking cathode at the bottom of the cell.[33,39] This arrangement ensured a lateral homogeneity of the deposition current density over the cathode area. Throughout the series, the Cu layer thickness was varied from 0.5 nm to 4.5 nm in steps of about 0.1 nm whereby the magnetic layer thickness was held constant at 2.7 nm (the actual values varied between 2.5 and 3.0 nm). The number of bilayer repeats was varied in a manner as to maintain a nearly constant total multilayer thickness of about 450 nm. The multilayers were deposited on Si(100)/Cr(5nm)/Cu(20nm) substrates whereby the adhesive Cr layer and the Cu seed layer were prepared by room-temperature evaporation on the Si wafer. More details of the sample preparation and characterization are described elsewhere.[47]

X-ray diffraction (XRD) technique was used to investigate the structure of the multilayer deposits. The structural results were described in detail in Ref. 47 and a brief summary is only given here. All the multilayers exhibited a predominantly fcc structure and a strong (111) texture along the growth direction. For small Cu layer thicknesses ($d_{Cu}$), a low amount of a hexagonal close-packed (hcp) phase of Co was revealed which phase practically disappeared at about $d_{Cu} = 2$ nm. On the other hand, no multilayer satellite reflections could be seen in this thickness range. For larger Cu layer thicknesses (2 nm < $d_{Cu}$ < 4 nm), clear satellite lines were visible due to the coherent superlattice structure of the multilayers. The bilayer repeat periods ($\Lambda = d_{Co} + d_{Cu}$) determined from the positions of the satellite reflections were in relatively good agreement with the nominal repeat periods, the experimental values being systematically larger by about 10 %. For multilayers with $d_{Cu} > 4$ nm, a degradation of the superlattice structure was indicated by the disappearance of satellite reflections. The good structural quality of multilayers in the range 2 nm < $d_{Cu}$ < 4 nm was also supported by the highest degree of texture and the least line broadening here.

The results of the structural study are consistent with a model according to which for $d_{Cu}$ < 2 nm there are pinholes in the Cu layers and these layers may also exhibit a large thickness fluctuation whereas there is a fairly perfect superlattice structure with continuous Cu layers for 2 nm < $d_{Cu}$ < 4 nm. The reason for the structural degradation for $d_{Cu} > 4$ nm may arise due to a change of the growth mode for thick Cu layers.[47]

*B. Measurement techniques*

The MR data were measured on 1 to 2 mm wide strips with the four-point-in-line method in magnetic fields between −8 kOe and +8 kOe in the field-in-plane/current-in-plane geometry at room temperature. Both the longitudinal (LMR, field parallel to current) and the transverse (TMR, field perpendicular to current) magnetoresistance components were measured. The following formula was used for calculating the magnetoresistance ratio: $\Delta R/R_0 = (R_H - R_0)/R_0$ where $R_H$ is the resistance in a magnetic field H and $R_0$ is the resistance value of the MR peak around zero field.

The room-temperature resistivity was determined in zero magnetic field by using a probe with four point contacts arranged along a line in fixed positions. A pure Cu foil of known thickness (ca. 25 μm) and having the same lateral dimensions as the multilayer sample to be measured was placed to a standard position in the probe. In this manner, a calibration constant of the probe was determined with the help of which, from the measured resistance of the sample with known thickness, the sample resistivity was determined.

The room temperature in-plane magnetization was measured in a vibrating sample magnetometer (VSM) throughout the whole Cu layer thickness range and in a SQUID magnetometer for two selected samples (one at low and one at high Cu layer thickness with AMR and GMR behaviour, respectively).

The electrical transport and VSM measurements were performed on the multilayers while being on their substrates. For the SQUID measurements, the multilayers were mechanically stripped off from the Si substrate. In order to see if the stripping has any influence on the magnetic properties, the M(H) loops were measured also for several samples after removing them from their susbtrate. The relative remanence remained the same as when measured on the substrates. The coercive field values changed by some 10 Oe but their evolution with Cu layer thickness was very similar to that obtained for multilayers on their substrates.

## III. RESULTS
### A. Zero-field electrical resistivity

The room-temperature electrical resistivity ($\rho_0$) in zero external magnetic field was determined for the present electrodeposited Co/Cu multilayers after the magnetoresistance measurements, i.e., after cycling the samples several times between –8 kOe and +8 kOe. Since the resistivity was measured for the approximately 450 nm thick multilayers on their substrate [Si/Cr(5nm)/Cu(20nm)], special care was taken of correcting for the shunting effect of the substrate metal layers. Therefore, by using the calibrated resistivity probe described in Section II.B, the resistivity was determined also for the Si/Cr(5nm)/Cu(20nm) substrate and $\rho_0 = 6.2$ μΩ·cm was obtained. The correctness of this substrate resistance value was checked by estimating the resistivity of the Cr(5nm)/Cu(20nm) substrate layer pair by applying a parallel resistor network model[48,49] for this bilayer. For bulk Cu metal, the room-temperature resistivity is $\rho_0(Cu) = 1.7$ μΩ·cm.[50] However, in thin films with a thickness smaller than the electron mean free path, the resistivity contribution due to surface scattering can be significant[51] and, therefore, the film resistivity can be much higher than the bulk value. Another contribution to the resistivity may come from grain boundary scattering since in thin films the lateral grain size is typically of the order of the film thickness.[48] Vancea and coworkers have reported in several papers[52] on the thickness dependence of the resistivity for evaporated thin Cu films. From these data, we can establish that at 20 nm thickness Cu films evaporated on a room temperature substrate, a condition corresponding to our case, exhibits a resistivity of $5 \pm 0.5$ μΩ·cm. If we use $\rho_{Cu}(20nm) = 5$ μΩ·cm and for the Cr(5nm) film the bulk value [$\rho_{Cr} = 12.9$ μΩ·cm (Ref. 50)], the resistivity of the Cr/Cu substrate bilayer is obtained as 5.7 μΩ·cm. Although we could not find data for the thickness dependence of Cr film resistivity, from the thickness dependencies reported for Cu and Nb films[52,53] we can infer that an increase of the Cr(5nm) film over the bulk value by a factor of 10 is reasonable. This leads us to the result $\rho_{Cr(5nm)/Cu(20nm)} = 6.2$ μΩ·cm, exactly the experimentally obtained value. Therefore, this value was used for correcting the experimentally determined resistivities for the Si/Cr(5nm)/Cu(20nm)//Co/Cu substrate/multilayer samples and the corrected values obtained for the Co/Cu multilayers are displayed in Fig. 1 (open circles). The correction due to substrate shunting effect amounts to about 1 μΩ·cm. The accuracy of the determination of the absolute value of the resistivity was estimated to be about 10 %. However, the relative accuracy of the resistivity measurement throughout the sample series investigated is significantly better, about 2 to 3 % only which is at most twice the data symbol size in Fig. 1.

As indicated by the solid trendline over the shunt-corrected data, the resistivity exhibits a maximum for Cu layer thicknesses around 1 nm. Our experimental results show a good qualitative agreement with the data of Lenczowski et al.[22] on electrodeposited Co/Cu multilayers: these authors have reported a similar resistivity decrease for Cu layer thicknesses from 1 nm to 6.5 nm although their resistivity values were systematically lower.

In a former work,[49] we investigated the thickness dependence of the resistivity in electrodeposited $Ni_{81}Cu_{19}$/Cu multilayers which was analyzed in terms of the parallel resistor model.[48,49] By using the known resistivities of bulk Cu metal and the bulk $Ni_{81}Cu_{19}$ alloy, it turned out from this analysis that whereas for large Cu layer thicknesses both the experimental data and the model values exhibited a decrease, the experimental resistivities were still much larger than the values from the parallel resistor model.

The situation is very similar for the present Co/Cu multilayers. Since the Cu-content is fairly low (0.6 at.% Cu) in the magnetic layers of our multilayers, for applying the parallel resistor model, we could take in principle the resistivity of electrodeposited bulk Co from an earlier work[54] according to which $\rho_0$ was found to be 10-15 μΩ·cm at room temperature.

However, the latter samples consisted of a mixture of hcp-Co and fcc-Co phases and also had a small grain size. By contrast, the present Co/Cu multilayers have an fcc structure and the lateral grain size is also definitely larger than in the previously studied electrodeposited Co since this is a prerequisite for the observation of a significant GMR. On the other hand, $\rho_0(300\ K) = 6\ \mu\Omega\cdot cm$ was reported for well-annealed, defect-free bulk hcp-Co by Laubitz et al.[55] These latter authors have also reported data[55] from which we can see that around the temperature of the hcp-fcc transition of bulk Co (at about 700 K) the resistivity of the fcc phase is by about 8 % smaller than that of the hcp phase. By assuming an identical temperature dependence of $\rho$ for both phases, we can assess $\rho_0(300\ K) = 5.5\ \mu\Omega\cdot cm$ for bulk fcc-Co. On the other hand, we can estimate an incremental resistivity of about 0.5 $\mu\Omega\cdot cm$ for the magnetic layer due to the small amount of Cu in it (this value is obtained under the plausible assumption that the resistivity increase due to alloyed Cu is the same for the Ni-Cu and the Co-Cu systems in their fcc phases and taking the incremental resistivity of Cu reported for fcc-Ni[49]). Thus, we end up with $\rho_0(300\ K) = 6\ \mu\Omega\cdot cm$ for the room-temperature resistivity of the bulk of the magnetic layer in the present Co/Cu multilayers. If we now apply the parallel resistor model[48,49] with value for the magnetic layer and with the bulk Cu resistivity, the dashed line in Fig. 1 indicates the resultant resistivity in this model, being well below the experimental data also for the Co/Cu multilayers.

As noted above, in thin films electron scattering events at the surfaces can dominate in the total resistivity[51] and, analogously, the same happens due to the interfaces in nanoscale multilayers. Therefore, the additional resistivity observed in both multilayer systems over the value from the parallel resistor model on the basis of bulk resistivities comes mainly from interface scattering. This contribution can be dominant over bulk-type scattering events within the magnetic layers if the layer thicknesses become comparable to the electron mean free path of the bulk form of the layer constituents. With increasing Cu layer thickness, the total resistivity should decrease since the interface scattering is reduced and the volume fraction of the low-resistivity Cu layer thickness increases. Even if there is a contribution from the increased number of grain boundaries in thin films, the dominant term originates from interface scattering.

With decreasing Cu layer thickness, the total resistivity will be more and more dominated by the interface scattering events and, therefore, $\rho_0$ should show an increase as actually observed down to about $d_{Cu} = 1$ nm (Fig. 1). On the other hand, the fall of $\rho_0$ for Cu layer thicknesses below 1 nm is an indication that the layered structure becomes destroyed since here the Cu layers may be no longer continuous. As a consequence, conduction electrons travelling between two adjacent Co layers can pass also through discontinuities of the Cu layer, i.e., travelling in Co only. In this sense, reduced continuity of the Cu layers results in more and more percolation of adjacent Co layers and conduction electrons tend to "feel" more and more a bulk-like Co environment. All this leads then to a diminution of the resistivity as observed for very thin Cu layers (Fig. 1). It is important to note that a linear extrapolation of the trendline over the experimental data (see dotted line in Fig. 1) to $d_{Cu} = 0$ yields very accurately the resistivity value assumed for bulk fcc-Co (6 $\mu\Omega\cdot cm$) which is a surprisingly good agreement with expectation. Although a few data points in Fig. 1 deviated markedly from the general trend (what may be due to a scatter from sample to sample rather than an experimental error associated with the resistivity determination), this observation provides further justification for the correctness of the decline of the trendline below 1 nm Cu layer thickness.

The above interpretation of the electrical resistivity data is in full conformity with the conclusions of the structural study[47] summarized in section II.A.

*B. Magnetoresistance*

The magnetoresistance behaviour of the present electrodeposited Co/Cu multilayers exhibits two distinct types as exemplified in Fig. 2. For multilayers with $d_{Cu} < 1.5$ nm, the LMR and TMR components have different signs, their difference at high fields defining the magnitude of AMR.[6] These samples exhibit a typical bulk FM type MR behaviour just as bulk Ni[6] or Co[54] metals. On the other hand, for multilayers with $d_{Cu} > 1.5$ nm both the LMR and TMR components were found to be negative and exhibited higher saturation values in comparison with samples showing AMR. This indicates a clear GMR behaviour for Cu layer thicknesses larger than 1.5 nm.

For both thickness ranges, the high-field region of the MR(H) curves were nearly linear above a saturation field $H_s$ of about 2 to 3 kOe (cf. Fig. 2). Following the procedure of Lenczowski et al.,[22] extrapolations from this linear region to H = 0 were considered as the saturation values of the corresponding magnetoresistance components as shown in Fig. 2. The linear decrease beyond the saturation field ($H_s$) is due to the gradual alignment of the magnetic moments with increasing magnetic field (so-called paraprocess) which results in a reduction of the electron scattering on thermally fluctuating atomic magnetic moments.[6] The MR(H) curve of the multilayer sample is shown on an enlarged scale in the inset of Fig. 2 where the MR peak position ($H_p$) is also defined. As we shall see later, the variation of $H_p$ correlates well with that of the coercive field $H_c$ although their magnitudes are not necessarily equal. This is because $H_c$ corresponds to the state with zero average magnetization of the whole sample whereas $H_p$ is the magnetic field value where the largest degree of antiparallel (AP) alignment of first-neighbor layer magnetizations is realized.

The measured saturation MR values are plotted in Fig. 3 as a function of the Cu layer thickness for both the LMR and TMR components. A fairly monotonous evolution can be established for almost the whole Cu thickness range. The bulk FM-type MR behaviour (AMR) prevails up to about 1.5 nm in which range the magnitudes of LMR > 0 and TMR < 0 components are nearly constant. A GMR behaviour (LMR < 0, TMR < 0) develops beyond about 1.5 nm Cu layer thickness. The GMR magnitude increases continuously with a maximum around 3.5 to 4 nm Cu layer thickness and slightly decreases thereafter. The vertical arrows in Fig. 3 indicate the approximate positions of the first three GMR maxima observed for sputtered fcc(111) Co/Cu multilayers.[5,56,57]

The clear absence of an oscillatory GMR behaviour can be established for the present electrodeposited Co/Cu multilayers. Especially, there are no distinct features in the GMR magnitude at the usual positions of the AF maxima in the oscillatory interlayer exchange coupling.[5]

It should be noted that the occurrence of an AMR behaviour for $d_{Cu} < 1.5$ nm can be well explained with the presence of pin-holes in the Cu layers, in agreement with the conclusions derived from our previous XRD measurements[47] and from the above described zero-field resistivity data for such thin Cu layers. On the other hand, the low saturation fields of the MR(H) curves and the linear MR(H) behaviour for $H > H_s$ in Fig. 2 clearly demonstrate that for $d_{Cu} > 1.5$ nm we have to account for GMR due to scattering events for electron paths between two FM layers, just as for the GMR of physically deposited FM/NM multilayers. This can only occur if in this Cu layer thickness range the FM layers are separated by a sufficiently thick and continuous non-magnetic spacer layer (at least over fairly large areas) that prevents a FM exchange coupling between the neighbouring magnetic layers. Again, the XRD[47] and zero-field resistivity data (Section III.A) give independent evidence for this microstructural model. The magnetic data to be presented in the next section provide further support for this picture. At the same time, they also allow us to better understand the evolution of microstructure, interlayer coupling and GMR magnitude with Cu layer thickness.

## C. Magnetic properties

For both low and high Cu layer thicknesses, FM type magnetization curves were obtained for the electrodeposited Co/Cu multilayers as demonstrated in Fig. 4 for two selected samples, one with AMR ($d_{Cu}$ = 0.9 nm) and one with GMR behaviour ($d_{Cu}$ = 3.0 nm). From a comparison of the low-field and high-field data, we could infer that the relative remanence $M_r/M_s$ is 0.74(1) and 0.76 for the two selected multilayers, respectively. This means that the remanence value of the multilayer with GMR behaviour practically equals the remanence of the sample with definitely FM coupling of the magnetic layers (AMR behaviour). It should be noted that very similar findings were reported by Lenczowski et al.[22] in that the relative remanence was reported to be between 0.7 and 0.8 for two electrodeposited Co/Cu multilayers, one with AMR and another one with GMR.

It can be concluded from these results that there is no AF coupling between the magnetic layers in the GMR multilayer since, then, the remanence would be significantly reduced with respect to the AMR multilayer. Along this line, we may say that the increase of GMR magnitude with increasing Cu layer thickness does not derive from an increase of the AF coupling but, instead, from a reduction of the FM coupling between the magnetic layers due to the improving perfectness of the separating spacer layers. In case of a reduced FM coupling, the adjacent layer magnetizations remain no longer parallel in zero magnetic field but they can align with respect to each other at various angles. With weakening FM coupling, this angle can increase and with increasing degree of disalignment, such FM layer pairs give rise to a larger and larger GMR contribution. The source of this disalignment for weakly FM-coupled or completely uncoupled FM layers may be several factors. The magnetization for such magnetic layers lies in the layer plane and if the domain wall energy determined by the exchange constant between magnetic atoms and the anisotropy energy is large, each magnetic layer may remain a single domain. The orientation within a plane is determined by the local anisotropy, mainly of magnetocrystalline origin. In an fcc crystal with the (111) lattice plane parallel with the layer plane as in the present Co/Cu multilayers, there are several equivalent easy axes within the layer plane. Reducing the magnetic field from saturation (fully aligned state) to zero, the magnetization of each layer falls into one of the possible two orientations of the numerous easy axes available and for a given layer this happens rather independently of the adjacent magnetic layers if their FM coupling is sufficiently weak or is completely absent. This may yield a rather random mutual orientation of the adjacent layer magnetization orientations, leading to a large GMR contribution. For uncoupled magnetic layers, if the domain wall energy is small, the magnetization of each layer may split into magnetic domains (such a situation is visualized in Ref. 58). In lack of an interlayer coupling, the magnetizations of opposing regions in adjacent layers have a great chance to be disaligned, again leading to a GMR effect.

The high-field magnetization curves displayed in Fig. 4 for magnetization values above remanence indicate a slight difference in the approach to saturation for the two multilayers. For very high fields, there is a residual magnetization increment due to the high-field susceptibility (paraprocess) typical for metallic ferromagnets (actually, this is very small, it amounts to about 0.02 $M_s$ only for the upper 30 kOe field range) and this is expected to be nearly the same for both multilayers. The difference appears in the intermediate magnetic field range (a few tens of kOe). The AMR multilayer with bulk-like magnetic behaviour approaches saturation faster and the obstacle against saturation may stem from surface roughness and various magnetic anisotropies. The slower approach to saturation in the GMR multilayer can probably be ascribed to the additional presence of SPM regions amounting to about 2 % of the total magnetic material as judged from the observed difference between the two multilayers. The occurrence of such a small SPM fraction in magnetic/non-magnetic multilayers is quite reasonable.

From the high-field SQUID measurements for the two selected multilayer samples, the saturation magnetization was determined. By taking into account the nominal layer thicknesses, 161 emu/g and 173 emu/g were obtained for the saturation magnetization of the magnetic layer of the multilayers with $d_{Cu} = 0.9$ nm (AMR behaviour) and $d_{Cu} = 3.0$ nm (GMR behaviour), respectively. The agreement with the saturation magnetization of pure Co metal (160 emu/g) is very good for the AMR multilayer and is within less than 10 % for the GMR multilayer (the poorer agreement in the latter case may partly come from the much smaller amount of magnetic material in this sample).

The coercive field ($H_c$) values obtained from the low-field hysteresis loops (see inset in Fig. 4) increased from about 20 Oe up to about 100 Oe with increasing Cu layer thickness as shown in Fig. 5 where also the $H_p$ values derived from the MR measurements are displayed. The evolution of the $H_p$ and $H_c$ data are in good agreement with each other. From the data, we can establish a kind of saturation at around 100 Oe just for the largest Cu layer thicknesses.

At low Cu layer thicknesses, the observed $H_c$ and $H_p$ values are in good agreement with what we have reported earlier[25,26] for similar multilayers. The lowest coercive field values obtained match well the data of Munford et al.[59] on thick (several 100 nm) Co films electrodeposited on Si substrates. These low coercivity values can already be considered as characteristic of bulk Co with predominantly fcc structure.

The coercivity results on the present multilayers can be understood in terms of the same structural features as already outlined above. The bulk-like $H_c$ and $H_p$ data observed for the lowest Cu layer thicknesses (Fig. 5) are a natural consequence of the percolation of Co layers via pin-holes in the Cu layers. With increasing Cu layer thickness, the magnetic layers become more and more perfectly separated due to the progressively improving continuity and/or uniformity of the Cu layers, reducing the strength of the FM coupling between adjacent Co layer magnetizations. This can also be expressed by saying that the "effective" thickness of the magnetic layers is somewhat higher than the actual one in case of a non-zero FM coupling between these layers which leads to a lower coercivity than would be expected for the actual magnetic layer thickness. As a result, with diminishing FM coupling between the magnetic layers upon increasing the Cu layer thickness, the multilayer coercive field should gradually increase to a value characteristic of individual, uncoupled Co layers with a thickness of about 2.7 nm. It is well-known that the $H_c$ of thin ferromagnetic layers increases roughly proportionally with the inverse of the layer thickness.[59,60]

Upon having put all this together, we can now return to Fig. 3 and try to explain the continuous increase of the GMR magnitude with $d_{Cu}$. Namely, as the Cu layer thickness increases from 1.5 nm to 3.5 nm, the degree of FM coupling between magnetic layers is reduced, the layers become more and more uncoupled, in zero field having their magnetization more and more randomly oriented with respect to each others and, especially, with respect to the adjacent layers. This is just what favors the occurrence of a larger and larger GMR effect as actually observed.

### D. Correlation between multilayer structure quality and GMR

We could see that all the experimental data (zero-field resistivity, magnetoresistance and coercivity) presented above in Section III for the current electrodeposited Co/Cu multilayer series are in conformity with the presence of pin-holes in thin Cu layers and a gradually improving continuity and/or thickness uniformity of the spacer layer with its increasing average thickness. Our previous structural study by XRD[47] on the same samples provided more direct evidence for such a structural model of these electrodeposited Co/Cu multilayers.

It should be pointed out at the same time that the different experimental methods suggest slightly different critical Cu layer thicknesses beyond which a significant decrease in the pin-hole density and/or an improvement in thickness homogeneity occurs. This is simply a consequence of the fact that each experimentally measured quantity probes differently the microstructure of multilayers under study. An important point is, however, that the Cu-thickness range with a clear GMR effect correlates well with the highest superlattice quality in the present multilayer series in terms of the presence of satellite reflections, the narrowest XRD lines and the strongest texture, pointing toward a large degree of structural perfectness.[47] Even the slight decrease in the GMR magnitude for $d_{Cu} > 4$ nm is unambiguously reflected[47] by the disappearance of the superlattice reflections, loss of texture degree and increase of amount of defects (e.g., decreasing grain size), all these features indicating structural degradation.

The presence of a small fraction of hcp-Co for Cu layer thicknesses below about 2 nm suggests[47] that there should be definitely pinholes here, enabling the growth of the equilibrium hcp-Co structure. The disappearance of the hcp-Co phase beyond 2 nm Cu thickness[47] indicates a considerable reduction of the pin-hole density. Here, a fluctuation of the Cu layer thickness may still prevail which can enable an FM coupling between adjacent magnetic layers over some areas but then the layer thickness uniformity gradually improves, finally not allowing an FM exchange coupling to occur through sufficiently thick Cu layers. The uncoupled magnetic layers can then develop a higher GMR due to their random relative orientation in zero magnetic field. For large thicknesses where the FM coupling completely disappears the GMR cannot increase anymore but saturates. After saturation, we observed a slight reduction of the GMR magnitude that may have occurred due to some structural degradation revealed by XRD[47] or simply because with increasing bilayer repeat period the interface density decreases and this should result in a reduction of the GMR magnitude.

## IV. DISCUSSION

In the following, we shall discuss only results on multilayers exhibiting a GMR behaviour (LMR < 0 and TMR < 0). First, a comparison will be made with former results on electrodeposited Co/Cu multilayers, by critically examining the reported spacer layer thickness dependencies. After coming to a conclusion about the absence of an oscillatory interlayer exchange coupling and GMR in these systems, we shall discuss why we can still observe a significant GMR in lack of AF coupling.

### A. Ferromagnetic and superparamagnetic contributions to the GMR in (electrodeposited) multilayers

Before performing a comparison and analysis of conflicting results concerning the GMR oscillation, we should first mention a specific feature of the GMR in electrodeposited multilayers. Namely, in many previous studies the magnetic field dependence of the magnetoresistance, MR(H), was found to be very different from that of physically deposited multilayers exhibiting clear AF coupling. In electrodeposited multilayers, the MR(H) curves were reported, especially at small spacer layer thicknesses, not to saturate up to magnetic fields beyond 10 kOe. By contrast, for physically deposited multilayers MR saturation against the AF-coupling can usually be achieved well below 10 kOe even at the first AF maximum (spacer thickness around 1 nm), whereas at the second and third AF maximum the saturation field is of the order of a few hundred oersted only.[5,57,61]

In order to understand the origin of such a difference, we consider first the classical magnetic/non-magnetic multilayers in which the magnetic layers contain FM regions only (most multilayers produced by physical deposition methods exhibit this behaviour). In such cases, the GMR effect arises from spin-dependent scattering originating from electron paths of the type "FM region 1 → NM region → FM region 2" and this is the conventional $GMR_{FM}$ term observed in physically deposited multilayers[1-5] which saturates at the above mentioned magnetic fields. However, it has been shown recently[43,44] that a non-saturating behaviour frequently observed in magnetic/non-magnetic multilayers produced by any methods can be successfully explained by the presence of SPM regions in the magnetic layers. An important consequence of the presence of SPM regions in multilayers is that there will be another contribution called $GMR_{SPM}$ which is due to spin-dependent scattering for electron paths "SPM region → NM region → FM region" (or in opposite order). The occurrence of electron paths "SPM region 1 → NM region → SPM region 2" was found to be negligible in the multilayer systems[44,46] as opposed to conventional granular metals.[62,63]

It has been found for electrodeposited Co-Cu/Cu multilayers with non-saturating MR(H) behaviour[44,46] that beyond technical saturation of the magnetization at about $H_s$ = 2 to 3 kOe, the field dependence of the magnetoresistance MR(H) can be described by the Langevin function L(x) where x = μH/kT with μ constituting the average magnetic moment of a SPM region. Beyond the saturation of ferromagnetic regions (H > $H_s$), the $GMR_{FM}$ and the AMR terms are saturated and, hence, their contributions remain constant for magnetic fields above $H_s$, apart from a small, linearly decreasing term (due to the paraprocess). Therefore, the contribution of the $GMR_{FM}$ and AMR terms cannot be separated from each other at H > $H_s$, and their sum will be denoted as a single $MR_{FM}$ term.

In this manner, one can describe the MR(H) data for magnetic fields H > $H_s$ in the form[44]

$$MR(H) = MR_{FM} + GMR_{SPM}\, L(x), \tag{1}$$

whereby $MR_{FM}$ = AMR + $GMR_{FM}$ is a constant term.

This decomposition method has been recently successfully applied to analyze the Co-layer thickness dependence of electrodeposited Co-Cu/Cu multilayers.[64] It should be noted that the occurrence of SPM regions is not restricted to electrodeposited magnetic/non-magnetic multilayers[25,26,28,33,39,41-46] since magnetic measurements revealed the presence of an SPM contribution to the magnetization also in multilayers prepared by physical deposition methods.[65-75] Specifically, the field dependence of the magnetoresistance in MBE-grown Co/Cu multilayers[72] could be well fitted by a Langevin function which implies the same magnetoresistance mechanisms as described above for the case of electrodeposited multilayers. The decomposition of GMR into FM and SPM contributions as suggested in Ref. 44 was also helpful in understanding the observed behaviour of sputtered Co/Cu[74] and Fe/In[76] multilayers.

Since the oscillatory GMR arising from an oscillatory exchange coupling of the layer magnetizations can be associated with the $GMR_{FM}$ term only, plotting the total GMR magnitude versus spacer layer thickness does not necessarily provide information on the true thickness dependence of the $GMR_{FM}$ contribution. This has been clearly demonstrated for a series of electrodeposited Co-Cu/Cu multilayers[28] for which the total GMR measured at a fixed magnetic field exhibited a minimum with increasing Cu layer thickness. After separating the $GMR_{FM}$ and $GMR_{SPM}$ terms, it turned out that the minimum was the result of an interplay between a decreasing $GMR_{SPM}$ term and an increasing $GMR_{FM}$ term. Therefore, when searching for an oscillatory GMR in electrodeposited multilayers, evidently the $GMR_{FM}$ term should be separated out from the total measured GMR or the $GMR_{SPM}$ term should be suppressed as much as possible by the preparation conditions.

As was shown in Section III.B, for the present Co/Cu multilayers the MR(H) curves became linear for magnetic fields above 2 to 3 kOe. This means that the SPM contribution was negligible in these samples, in agreement with the conclusions of magnetic measurements (see Section III.C) and the $MR_s$ values established by extrapolation to H = 0 can be identified as the $MR_{FM}$ = AMR + $GMR_{FM}$ term. Therefore, these data can be considered as being characteristic of an MR contribution due to spin-dependent scattering events between FM parts of the magnetic layers in our multilayers. In this sense, the saturation magnetoresistance data on the present electrodeposited Co/Cu multilayers with $d_{Cu}$ > 1.5 nm, apart from a small AMR contribution (typically not more than 0.5 %), correspond to the conventional GMR observed in physically deposited FM/NM multilayers. An average of the saturation values of the longitudinal and transverse MR components ($LMR_s$ and $TMR_s$, respectively) according to the formula $GMR_s$ = (1/3) $LMR_s$ + (2/3) $TMR_s$ which takes care for a correction due to the AMR can be finally identified as the $GMR_{FM}$ term of our multilayers and these data are shown in Fig. 6 (open triangles) as a function of the Cu layer thickness $d_{Cu}$.

*B. Comparison with former results on the spacer layer thickness dependence of GMR in electrodeposited Co/Cu multilayers*

1. Reports without oscillatory GMR behaviour

For comparison with the present results, we have first selected those reports where it can be established that the GMR data correspond to the $GMR_{FM}$ term similarly as discussed for our samples above. This was the case with our two previous works[26,28] and with the results of Lenczowski et al.[22] and Li et al.[27] and all these former data are also displayed in Fig. 6. Although the magnitude of GMR varies from study to study (probably due to differences in actual layer thicknesses, preferred texture, substrate material and other details of the electrodeposition process), the general trend is that (i) a clear GMR effect develops above a certain Cu layer thickness of about 1 nm only, (ii) the GMR magnitude increases monotonically with $d_{Cu}$ and (iii) a saturation or maximum occurs for Cu layer thicknesses around and above about 4 nm. A few further data of Lenczowski et al.[22] and Liu et al.[26] omitted from Fig. 6 show a qualitatively similar behaviour.

A monotonic GMR increase was observed also by Shima et al.[23] up to about 5 nm Cu layer thickness but their MR(H) curves indicate that saturation has not been achieved up to the maximum magnetic field applied and, therefore, their GMR values may contain an SPM contribution as well. The GMR results of Kainuma et al.[18] and Myung et al.[24] also reveal a maximum-like behaviour in the same range as for the data displayed in Fig. 6 but these authors have not shown MR(H) curves at all and in this manner we have no information on the eventual importance of a $GMR_{SPM}$ term.

The general conclusion is that the $GMR_{FM}$ term in electrodeposited Co/Cu multilayers does not exhibit an oscillatory behaviour as a function of the Cu layer thickness in those cases where we can unambiguously identify the appropriate $GMR_{FM}$ contribution characteristic for FM/NM multilayers.

2. Reports with "oscillatory" GMR behaviour

From among the papers reporting on an "oscillation" of the GMR magnitude for electrodeposited Co/Cu multilayers,[11-15] we discuss first the results by Jyoko et al.[13] For a multilayer with $d_{Cu}$ around 1 nm, these authors present an MR(H) curve which unambiguously reveals a dominant SPM contribution. Therefore, the high GMR value at this Cu layer thickness cannot originate from an AF coupled state and this conclusion is further supported by the M(H) curve reported for the same sample since it shows a large remanence whereas an AF-coupled state should exhibit a low remanence. For higher $d_{Cu}$ values, their multilayers display the typical MR(H) curves as observed also by us (split MR peaks, low saturation field) and have a similar evolution of the GMR magnitude as shown in Fig. 6. The results of Ueda et al.[12] show the same features: the non-saturating MR(H) curves obtained for Cu thicknesses around their first observed GMR maximum ($d_{Cu}$ ≈ 1.5 nm) are dominated by an SPM term whereas split MR peaks with low saturation fields appear for Cu layer thicknesses around 3 to 4 nm. It should also be noted that their first GMR maximum appears roughly at a Cu layer thickness where usually FM coupling is observed for physically deposited Co/Cu multilayers. Actually, their Cu layer thicknesses are probably even higher than the specified values as a consequence of the applied galvanostatic deposition technique due to a significant exchange reaction during the Cu deposition pulse as was pointed out, e.g., in Refs. 39 and 41.

Based on the argumentation presented above, we can conclude that in the above described two works[12,13] no evidence for a GMR oscillation corresponding to that observed in physically deposited Co/Cu multilayers can be identified.

The first observation of GMR oscillation in electrodeposited multilayers was reported by Bird and Schesinger[11] who even fitted their "oscillatory" GMR data for Co/Cu and Ni/Cu by an RKKY function. However, no details including MR(H) curves were presented in that short communication. Furthermore, the GMR magnitude was reported to be as high as in the corresponding sputtered counterparts which results have, however, never been reproduced by other researchers. For this reason, we have to treat these findings with appropriate caution.

There are still two further reports[14,15] which claim to have observed GMR oscillations in electrodeposited Co/Cu multilayers. However, in lack of sufficient details about sample preparation and magnetoresistance measurements, we cannot conclude about the reliability of these data.

*C. Origin of GMR in electrodeposited Co/Cu multilayers and explanation of observed GMR evolution with Cu layer thickness*

It was already discussed above that electrodeposited Co/Cu multilayers often exhibit a significant SPM fraction. This can easily occur if the magnetic layer itself is fairly thin (below about 1 nm). The thickness of the magnetic layers can also be substantially reduced, especially locally, under non-optimized deposition conditions[42] because of the dissolution of the magnetic (i.e., less noble) metal. In either case, small regions may become decoupled from the FM layers which then constitute SPM entities. We have also observed[25] that even if no Co dissolution of the deposited Co layer is expected to occur, under certain circumstances a low value (of about 1 nm or less) of both the Co and the Cu layer thicknesses results in a fairly large SPM fraction. A model by Ishiji and Hashizume[69] explains how a rough substrate can also lead to the development of SPM regions even in sputtered multilayers.

As mentioned above, it could be shown[44] that in the presence of SPM regions in a magnetic/non-magnetic multilayer, the magnetoresistance exhibits a strongly non-saturating character and its field dependence can be described by a Langevin function for magnetic fields above about 2 to 3 kOe. Even if the ratio of the SPM/FM volume fractions of the magnetic layers as deduced from magnetization measurements is as low as 0.1, the total observed GMR can be dominated by electron scattering events along electron paths between a FM and an SPM entity (the other GMR contribution is due electron scattering events for electron paths between two FM regions, being the sole contribution in multilayers with fully FM magnetic layers).

It can be established that in several previous works on electrodeposited multilayers this SPM type GMR contribution was the dominant term for low Cu layer thicknesses just around the value where the first GMR maximum was found to occur in sputtered Co/Cu multilayers. Evidently, this contribution cannot originate from an AF coupling and this is further supported by the usually much larger saturation fields as well.

If the fraction of the SPM regions is fairly low (at most a few percent of the total magnetic material), then the magnetic and magnetotransport behaviour of the ferromagnetic/non-magnetic multilayer system will depend on whether the spacer layer material is continuous or it contains a high density of pin-holes. In the following, we shall discuss electrodeposited Co/Cu multilayers with a negligible SPM fraction only. Below a thickness of typically 1 nm, the Cu spacer layers in electrodeposited multilayers usually contain a large density of pin-holes which provide a direct FM exchange coupling between adjacent layer magnetizations. In such a case, bulk-like FM behaviour with an AMR effect occurs due to a percolation of the magnetic layers via the pin-holes in the Cu layers. The same effect is observed if the Cu layer thickness fluctuation is significant and at some regions the very small spacer thickness enables a direct FM exchange coupling. With increasing average thickness, the continuity of the Cu layers increases and the reduced density of pin-holes as well as the improvement of Cu layer thickness uniformity weakens the FM exchange coupling between the magnetic layers which, thus, become gradually uncoupled. The uncoupled layer magnetizations will be randomly aligned and electron transitions between non-aligned adjacent layers can yield a larger and larger GMR effect as observed. At sufficiently large Cu layer thicknesses (around 3 to 4 nm), when the magnetic layers become completely uncoupled, there is no more increase in the randomness of the magnetization alignments and the GMR reaches saturation, in parallel with the saturation of the coercive force. The value of the latter quantity becomes then characteristic for thin individual magnetic layers. Since the relative remanence of the AMR and GMR multilayers was found to be practically the same, we have to conclude the absence of a significant AF coupling in the latter ones.

Beyond a certain spacer layer thickness, we have to expect a reduction of the GMR due to a decrease of the number of the magnetic/non-magnetic interfaces per unit thickness (dilution effect). A decrease of GMR is also expected when exceeding the Cu layer thickness through which the spin-memory is no longer preserved for the conduction electrons since then another pre-requisite for the observation of the GMR is not fulfilled.

It should be noted that Shima et al.[23] suggested a surface roughness model for electrodeposited Co/Cu multilayers in order to explain the absence of oscillatory behaviour. In this model, the authors have assumed that a Néel-type "orange-peel" coupling of magnetostatic origin provides a strong enough FM coupling to overcome the existing AF exchange coupling between adjacent magnetic layers. Even if this mechanism can explain the absence of GMR maxima at the expected positions of the AF coupling, in the intermediate Cu thickness regions the addition of the magnetostatic FM coupling to the FM exchange coupling would then provide a strong resulting coupling, i.e., a diminished GMR. By contrast, the $GMR_{FM}$ data displayed in Fig. 6 from various previous reports and from the present study show a uniquely monotonous increase of the GMR magnitude (at least up to the maximum beyond 3 nm Cu thickness), thus questioning the validity of the model of Shima et al.[23]

The explanation we proposed for explaining the evolution of the magnetoresistance in electrodeposited Co/Cu multilayers with spacer layer thickness is not restricted to systems produced by this method. Parkin and coworkers used very similar arguments for explaining the occurrence of GMR in lack of AF coupling[58] and the reduction of GMR for very thick Cu layers[77] in sputtered Co/Cu multilayers or the continuous increase of the GMR with Cu layer thickness in some MBE-grown Co/Cu multilayers.[58] Also, the SPM type GMR contribution as discussed above is not unique to electrodeposited multilayers since it was found in several physically deposited multilayers as well.[66,72-76]

## V. SUMMARY

In order to clarify the controversial results for the spacer layer thickness dependence of GMR in electrodeposited multilayers, a detailed study of the GMR evolution was performed on Co/Cu multilayers prepared under controlled electrochemical conditions with Cu layer thicknesses ranging from 0.5 nm to 4.5 nm. It turned out that for thin Cu layers (up to 1.5 nm) AMR only occurs. This could be explained by a high density of pin-holes in the thin spacer layers that enables the percolation of the magnetic layers yielding an overall bulk ferromagnetic (FM) like behaviour manifested in the observed AMR. For thicker Cu layers, a clear GMR was observed the magnitude of which increased up to a maximum at about 3.5 to 4 nm and with a slight decrease afterwards. The results of coercive field and zero-field resistivity measurements also indicated a transition from Cu layers with a high density of pin-holes to Cu layers with much better continuity and/or thickness uniformity at comparable thicknesses as deduced from the magnetoresistance data. A structural study reported earlier on the same multilayers[47] gave independent evidence for the microstructural features established here. According to magnetic measurements up to 50 kOe, the relative remanence for an AMR and a GMR multilayer was practically the same, hinting at the absence of an AF coupling between the magnetic layers. From an analysis of the present results and previously reported studies, it could be concluded that no well-documented evidence of an oscillatory GMR exists for electrodeposited Co/Cu multilayers. It was pointed out that the large GMR reported previously on such systems at Cu layer thicknesses around 1 nm can be well explained by the presence of a fairly large SPM fraction rather than being due to a strong AF coupling. In the absence of SPM regions, AMR prevails at low spacer thicknesses due to the dominating FM coupling via pin-holes as in the present case and in Ref. 22. With increasing continuity and thickness uniformity of the thicker and thicker spacer layers, the FM coupling strength is gradually reduced and finally disappears. This results in completely uncoupled magnetic layers with random magnetization orientations. As the magnetic layers become more and more randomly aligned with diminishing FM coupling, electron transitions between them provide an increasing GMR effect for larger spacer layer thicknesses.

As a main conclusion, we believe to have provided evidence that the absence of oscillatory GMR in electrodeposited multilayers is (i) partly due to the microstructural features revealed in the present work and by our former XRD study[47] which features result in an FM coupling for a very large range of spacer layer thicknesses and (ii) partly due to the absence of a significant AF coupling between the adjacent layers at the appropriate layer thicknesses.

Nevertheless, we still owe an explanation for the origin of the absence of a sizeable AF coupling between the adjacent magnetic layers in electrodeposited multilayers. Understanding this deficiency remains a great challenge and definitely requires studies of finer details of the microstructure such as interface roughness and intermixing which are known to be deleterious for the AF coupling. A strong reduction of the AF coupling was observed also in sputtered Co/Cu multilayers[78] due to residual gas impurities in the sputtering chamber and this may provide further hints in which direction to attempt an improvement of the multilayer electrodeposition technology.

It was also pointed out that the critical Cu layer thickness of the AMR to GMR transition beyond which the pin-hole density and/or layer thickness fluctuations are significantly reduced, varies from study to study. It is yet to be explored which electrodeposition parameters have a decisive influence in this respect. A progress in this field definitely requires further work on understanding the atomistic aspects of nucleation and layer growth during the electrodeposition process. There is certainly room for studying the influence of bath composition on the critical Cu thickness and to find eventually some surfactants with some beneficial effects as was the case with Pb in the growth of Co/Cu multilayers by MBE.[79]

**Acknowledgements** This work was supported by the Hungarian Scientific Research Fund (OTKA) through grants K 60821.

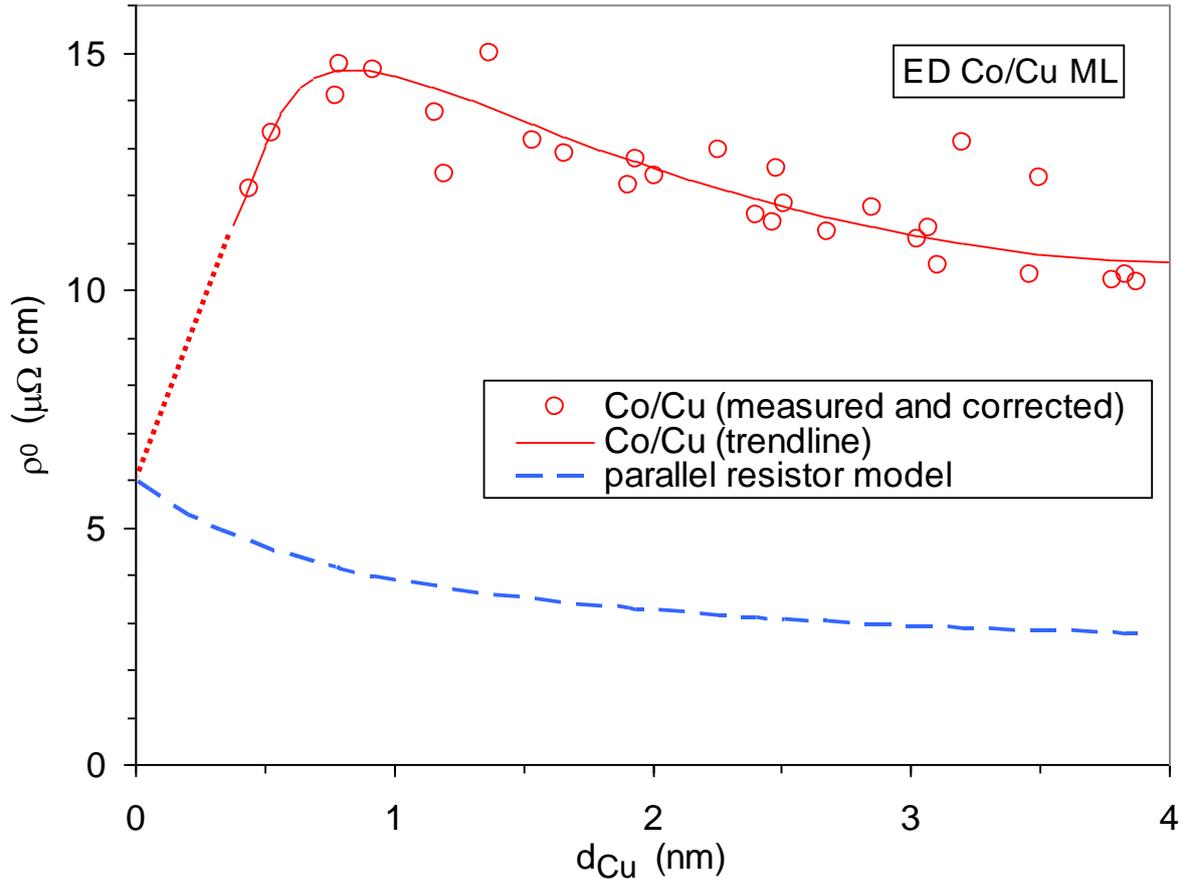

Fig. 1 (Color online) Room-temperature electrical resistivity ($\rho_0$) of electrodeposited Co/Cu multilayers in zero external magnetic field as a function of the Cu layer thickness with constant magnetic layer thickness $d_{Co} \approx 2.7$ nm. The symbols ○ represent experimental data after correction for the shunting effect of the Cr(5nm)/Cu(20nm) metallic substrate layers (see text for details). The error bar for each data point is at most twice the size of the data symbol. The solid line through the corrected experimental data indicates a trendline only. The dashed line represents the resistivity of a Co/Cu multilayer in a simple parallel resistor model,[48,49] calculated with bulk resistivity values of the individual layers. The dotted line is just a linear extrapolation of the experimental data to $d_{Cu} = 0$.

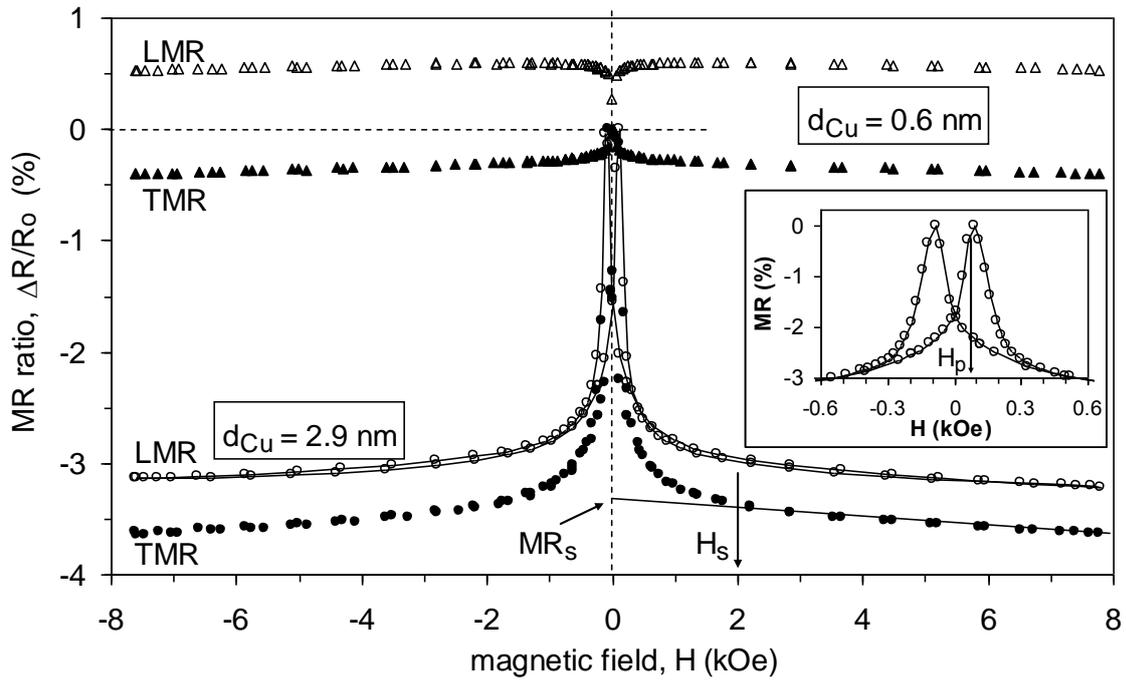

Fig. 2　Longitudinal (L, open symbols) and transverse (T, full symbols) components of the magnetoresistance (MR) for two electrodeposited Co/Cu multilayers: one exhibiting AMR (triangles) and one exhibiting GMR (circles). The spacer layer thickness is also indicated for both samples. The saturation field ($H_s$) is defined as the magnetic field above which the MR(H) variation can be considered as nearly linear. An extrapolation to H = 0 yields the saturation magnetoresistance ($MR_s$). The inset shows the MR(H) curve for the multilayer exhibiting GMR where the definition of $H_p$, the MR(H) peak position, is also given.

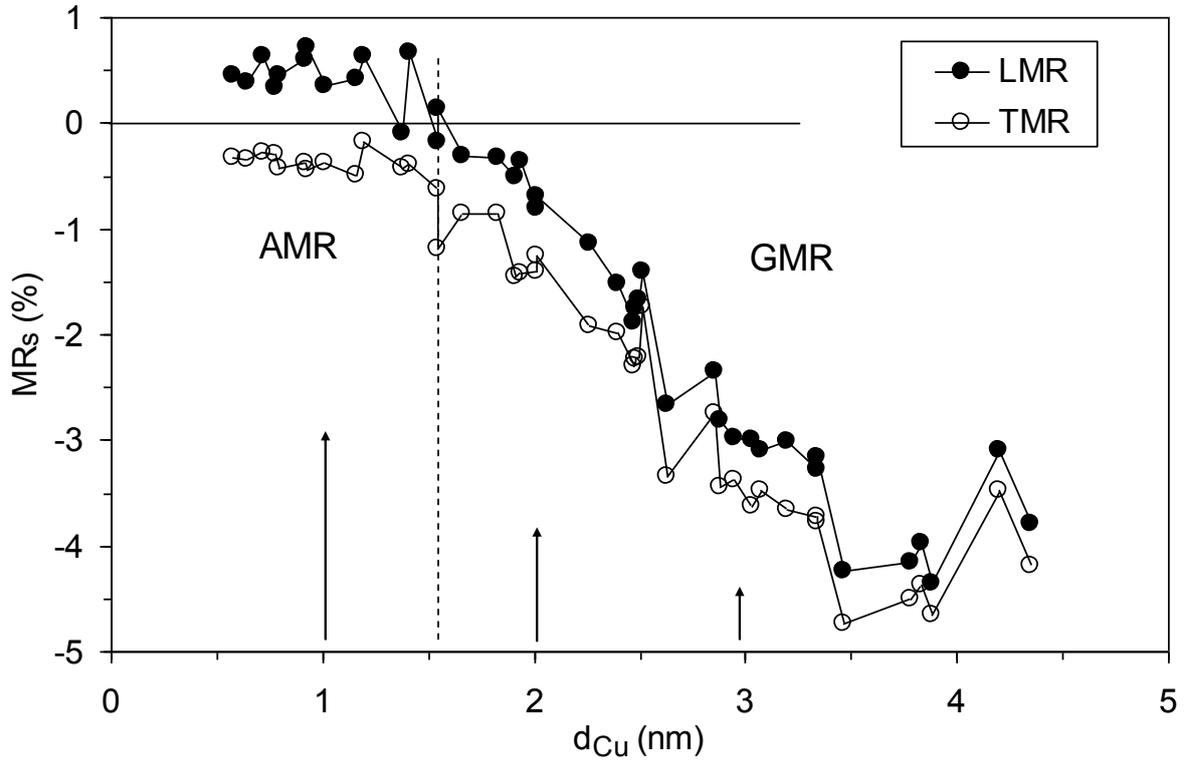

Fig. 3  Evolution of the longitudinal (LMR) and transverse (TMR) saturation components of the magnetoresistance MR for the investigated electrodeposited Co/Cu multilayers as a function of the Cu layer thickness $d_{Cu}$. For multilayers with $d_{Cu}$ not exceeding about 1.5 nm, the observed magnetoresistance is of the AMR type (LMR > 0; TMR < 0); for larger Cu layer thicknesses, the total observed magnetoresistance is dominated by GMR (LMR < 0; TMR < 0). The vertical dashed line separates the AMR and GMR thickness ranges. The vertical arrows denote the approximate positions of the GMR maxima reported for fcc(111) Co/Cu multilayers prepared by physical deposition methods.[5,56,57]

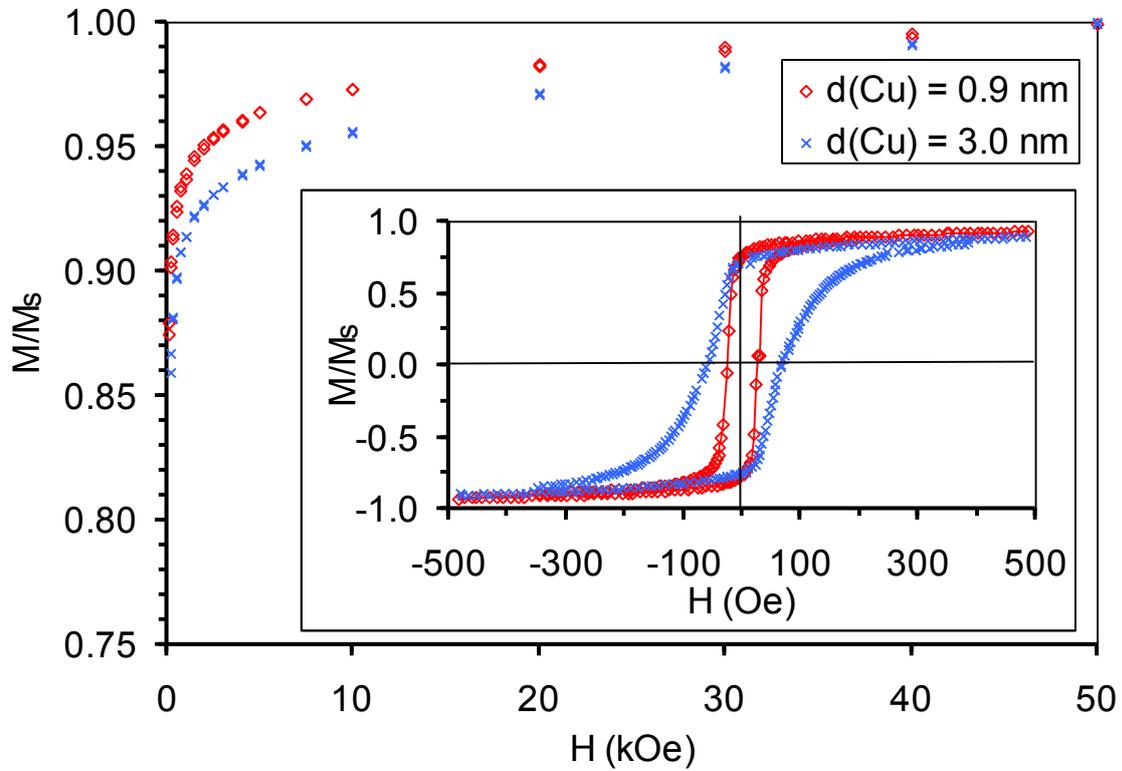

Fig. 4 (Color online) High-field magnetization curves normalized with the values measured at H = 50 kOe and displayed above the remanence values ($M_r/M_s \approx 0.75$) for electrodeposited Co/Cu multilayers with AMR behaviour ($d_{Cu}$ = 0.9 nm) and with GMR behaviour ($d_{Cu}$ = 3.0 nm). The inset shows the corresponding low-field magnetization curves with coercive field ($H_c$) values of 25 Oe and 63 Oe, respectively.

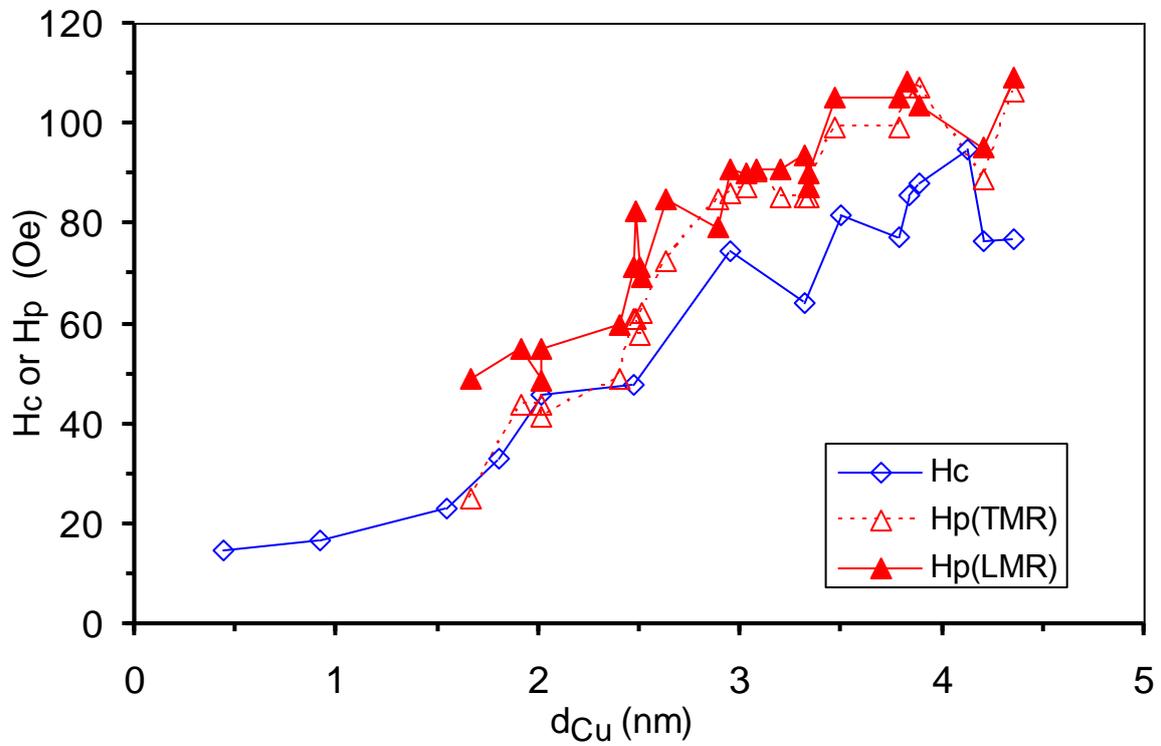

Fig. 5 (Color online) Magnetoresistance peak position value $H_p$ and coercive field $H_c$ for the investigated electrodeposited Co/Cu multilayers as a function of the Cu layer thickness $d_{Cu}$.

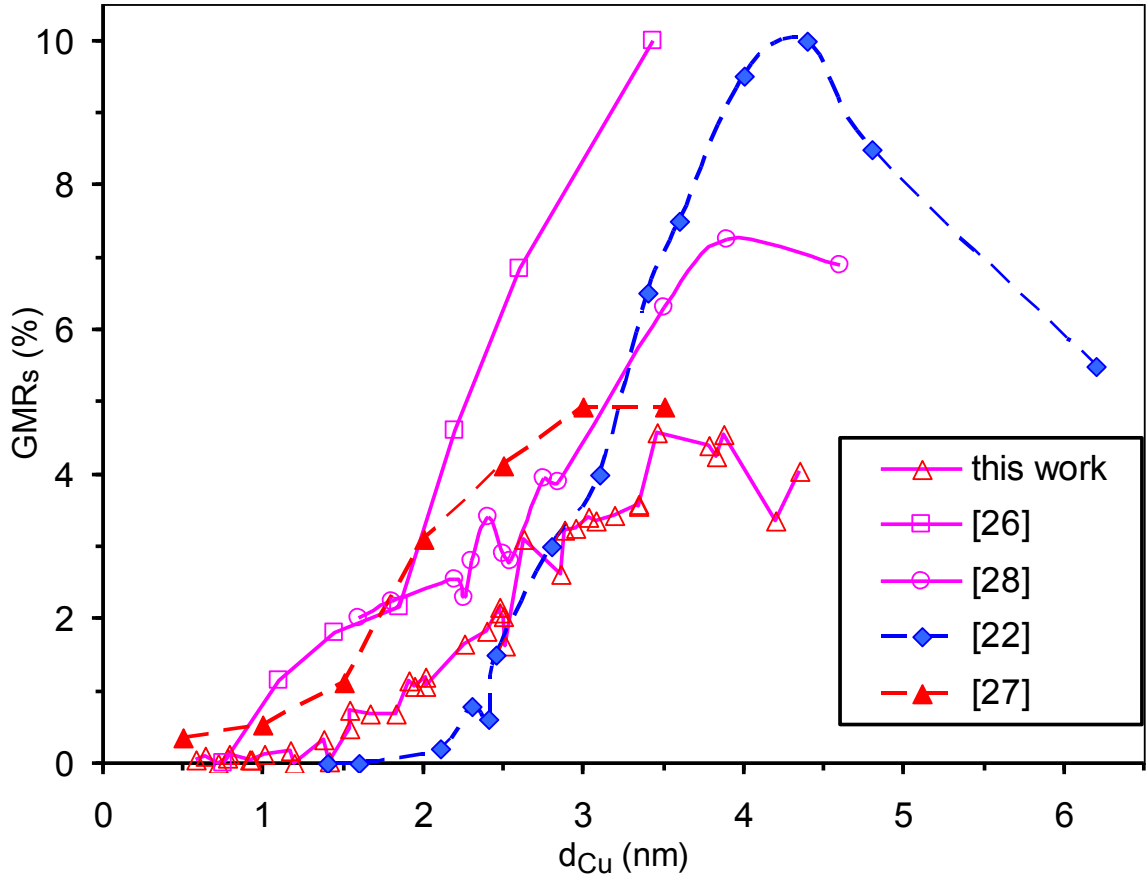

Fig. 6 (Color online) Evolution of the saturation GMR (GMR$_s$) in electrodeposited Co/Cu multilayers with Cu layer thickness d$_{Cu}$. The figures in [] indicate literature data references. It is noted that, according to a study reported in Ref. 41, in electrodeposited Co/Cu multilayers obtained under galvanostatic control, the actual layer thicknesses for the samples from Ref. 27 can be by as much as 1 nm higher than the original nominal thicknesses specified in that work.